\input{aipcheck}

\documentclass[
    ,final            
,numberedheadings 
  ]
  {aipproc}

\layoutstyle{6x9}

\def\la{\mathrel{\hbox{\rlap{\hbox{\lower4pt\hbox{$\sim$}}}\hbox{$<$}}}}
\def\ga{\mathrel{\hbox{\rlap{\hbox{\lower4pt\hbox{$\sim$}}}\hbox{$>$}}}}

\begin{document}

\title{Chandra Observations of Supernova 1987A}


\author{Sangwook Park}{
  address={Department of Astronomy and Astrophysics, 525 Davey Lab., Pennsylvania State University, University Park, PA 16802, USA}
}

\author{David N. Burrows}{
  address={Department of Astronomy and Astrophysics, 525 Davey Lab., Pennsylvania State University, University Park, PA 16802, USA}
}

\author{Gordon P. Garmire}{
  address={Department of Astronomy and Astrophysics, 525 Davey Lab., Pennsylvania State University, University Park, PA 16802, USA}
}

\author{Richard McCray}{
  address={JILA, University of Colorado, Box 440, Boulder, CO 80309, USA}
}

\author{Judith L. Racusin}{
  address={Department of Astronomy and Astrophysics, 525 Davey Lab., Pennsylvania State University, University Park, PA 16802, USA}
}

\author{Svetozar A. Zhekov}{
  address={Space Research Institute, Moskovska Strasse 6, Sofia 1000, Bulgaria}
}

\begin{abstract}
We have been monitoring Supernova (SN) 1987A with {\it Chandra X-Ray 
Observatory} since 1999. We present a review of previous results from 
our {\it Chandra} observations, and some preliminary results from new 
{\it Chandra} data obtained in 2006 and 2007. High resolution imaging 
and spectroscopic studies of SN 1987A with {\it Chandra} reveal that 
X-ray emission of SN 1987A originates from the hot gas heated by 
interaction of the blast wave with the ring-like dense circumstellar 
medium (CSM) that was produced by the massive progenitor's equatorial 
stellar winds before the SN explosion. The blast wave is now sweeping 
through dense CSM all around the inner ring, and thus SN 1987A is 
rapidly brightening in soft X-rays. At the age of 20 yr (as of 2007 
January), X-ray luminosity of SN 1987A is $L_{\rm X}$ $\sim$ 2.4 
$\times$ 10$^{36}$ ergs s$^{-1}$ in the 0.5$-$10 keV band. X-ray 
emission is described by two-component plane shock model with electron 
temperatures of $kT$ $\sim$ 0.3 and 2 keV. As the shock front interacts 
with dense CSM all around the inner ring, the X-ray remnant is now 
expanding at a much slower rate of $v$ $\sim$ 1400 km s$^{-1}$ than 
it was until 2004 ($v$ $\sim$ 6000 km s$^{-1}$).

%
%
\end{abstract}

\maketitle


\section{Introduction}

Supernova (SN) 1987A, the nearest SN in four centuries, occurred
in the Large Margellanic Cloud (LMC). The identification of a Type II SN 
from a blue supergiant progenitor and the detection of neutrino bursts 
associated with the SN indicate a core-collapse explosion of a massive 
star \cite{arnett89}. SN 1987A, providing these fundamental parameters 
and being located at a near distance ($d$ = 50 kpc), is a unique 
opportunity for the study of a massive star's death and the subsequent 
birth of a supernova remnant (SNR) in unprecedented detail. 

About 10 yr after the SN explosion, the blast 
wave started to interact with the ``inner ring'' of dense circumstellar 
medium (CSM) \cite{michael00}, which is believed to be produced by the 
equatorial stellar winds of the massive progenitor star. This shock-CSM 
interaction resulted in a dramatic brightening of SN 1987A in soft 
X-rays, which provides an excellent laboratory for the X-ray study 
of the evolution of an optically thin thermal plasma in nonequilibrium 
ionization (NEI) as the shock propagates through a complex density 
gradient of the dense CSM. As the rapidly brightening X-rays begin to 
illuminate the interior of the SN, metal-rich ejecta expelled 
from the massive star's core will begin to glow optically, allowing us 
to study SN nucleosynthesis yields.  Then, a few decades from now, 
when these newly-formed elements begin to cross the reverse shock surface, 
we will be able to measure the distribution of these elements in more 
detail through their X-ray emission.
Neutrino bursts were a strong support for a core-collapse explosion, 
and thus for the creation of a neutron star which should become bright 
in X-rays. 

High resolution imaging and spectroscopic studies of SN 1987A with 
{\it Chandra X-Ray Observatory} are an ideal tool for the X-ray study 
of SN 1987A. We have thus been observing SN 1987A with {\it Chandra} 
since its launch in 1999, roughly twice a year, in order to monitor the 
earliest stages of the evolution of the X-ray remnant of SN 1987A. 
We here review previous results from our {\it Chandra} observations of 
SNR 1987A \cite{burrows00,michael02,park02,park04,park05,park06,zhekov05,
zhekov06}, and present some preliminary results from the latest {\it 
Chandra} observations. 

\section{Observations}

Our {\it Chandra} observations of SNR 1987A are listed in Table~1.
As of 2007 January, we have performed a total of sixteen {\it
Chandra} observations of SNR 1987A, including two deep gratings
observations. Data reduction and analysis process have been
described in the literatures \cite{burrows00,michael02,park04,zhekov05}.

\begin{table}
\begin{tabular}{ccccc}
\hline
\tablehead{1}{c}{b}{Observation ID}
& \tablehead{1}{c}{b}{Date\\(Age)\tablenote{Days since SN}}
& \tablehead{1}{c}{b}{Instrument\\(Subarray)}
& \tablehead{1}{c}{b}{Exp.\\(ks)}
& \tablehead{1}{c}{b}{Counts}  \\
\hline
124+1387\tablenote{These observations were splitted by multiple sequences 
which were combined for the analysis.} & 1999-10-6 (4609) & ACIS-S+HETG
& 116.1 & 690\tablenote{Photon statistics are from the zeroth-order data.}\\
122 & 2000-1-17 (4711) & ACIS-S3 (None) & 8.6 & 607 \\
1967 & 2000-12-07 (5038) & ACIS-S3 (None) & 98.8 & 9030 \\
1044 & 2001-4-25 (5176) & ACIS-S3 (None) & 17.8 & 1800 \\
2831 & 2001-12-12 (5407) & ACIS-S3 (None) & 49.4 & 6226 \\
2832 & 2002-5-15 (5561) & ACIS-S3 (None) & 44.3 & 6427 \\
3829 & 2002-12-31 (5791) & ACIS-S3 (None) & 49.0 & 9277 \\
3830 & 2003-7-8 (5980) & ACIS-S3 (None) & 45.3 & 9668 \\
4614 & 2004-1-2 (6157) & ACIS-S3 (None) & 46.5 & 11856 \\
4615 & 2004-7-22 (6359) & ACIS-S3 (1/2) & 48.8 & 17979 \\
4640+4641+5362+5363+6099$^{\dagger}$ & 2004-8-26$\sim$9-5 ($\sim$6400) & 
ACIS-S+LETG & 289.0 & 16557$^{**}$ \\
5579+6178$^{\dagger}$ & 2005-1-12 (6533) & ACIS-S3 (1/8) & 48.3 & 24939 \\
5580+6345$^{\dagger}$ & 2005-7-14 (6716) & ACIS-S3 (1/8) & 44.1 & 27048 \\
6668 & 2006-1-28 (6914) & ACIS-S3 (1/8) & 42.3 & 30940 \\
6669 & 2006-7-28 (7095) & ACIS-S3 (1/8) & 36.4 & 30870 \\
7636 & 2007-1-19 (7271) & ACIS-S3 (1/8) & 33.5 & 32798 \\
\hline
\end{tabular}
\caption{Chandra Observations of SNR 1987A}
\label{tab:a}
\end{table}







\section{X-Ray Images}

\begin{figure}
  \includegraphics[height=0.68\textheight, angle=0]{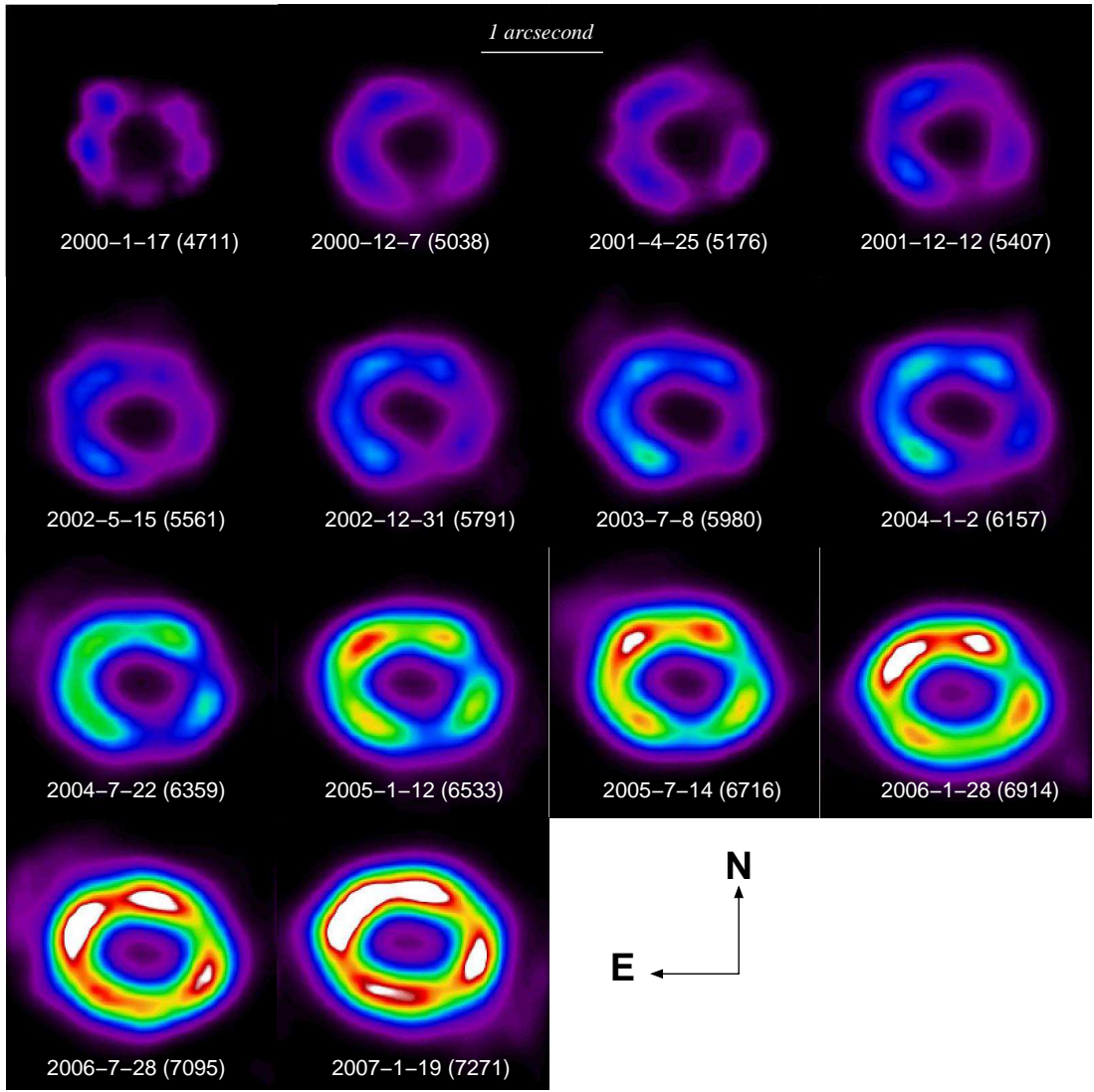}
  \caption{{\it Chandra} ACIS (in the 0.3$-$8 keV band) false-color images 
of SNR 1987A.  In each panel, the observation date and age (days since 
the SN, in parentheses) are presented.}
\end{figure}

Broadband {\it Chandra} ACIS images of SNR 1987A are presented in 
Fig.~1. We applied a subpixel resolution method \cite{tsunemi01}, 
deconvolved images with the detector point spread function (PSF), 
and then smoothed. The ring-like overall morphology of the X-ray remnant 
is evident. SNR 1987A has been brightening and expanding for the last
7 yr. Initially, the eastern side was brighter, but then the western 
side began brightening in early 2004 (day $\sim$6200). SNR 1987A is 
now bright all around the ring. Early images showed that the soft X-ray 
band images ($E$ $<$ 1.2 keV) were correlated with the optical images 
while the hard band ($E$ $>$ 1.2 keV) image matched the radio images 
\cite{park02}. These differential X-ray morphologies supported our 
interpretation that soft X-rays are produced by the decelerated shock 
entering dense protrusions of the inner ring and that hard X-rays originate
from the fast shock propagating through less dense regions between 
protrusions. Recent data show that the X-ray morphology is now nearly 
identical between the hard and soft bands, which is perhaps expected 
as an increasing fraction of the blast wave shock front is reaching 
dense CSM all around the inner ring \cite{park04}. The 0.3$-$8 keV band 
count rate is now $\sim$0.98 c s$^{-1}$, which is $\sim$14 times brighter 
than it was in 2000.

\begin{figure}
  \includegraphics[height=0.63\textheight, angle=-90]{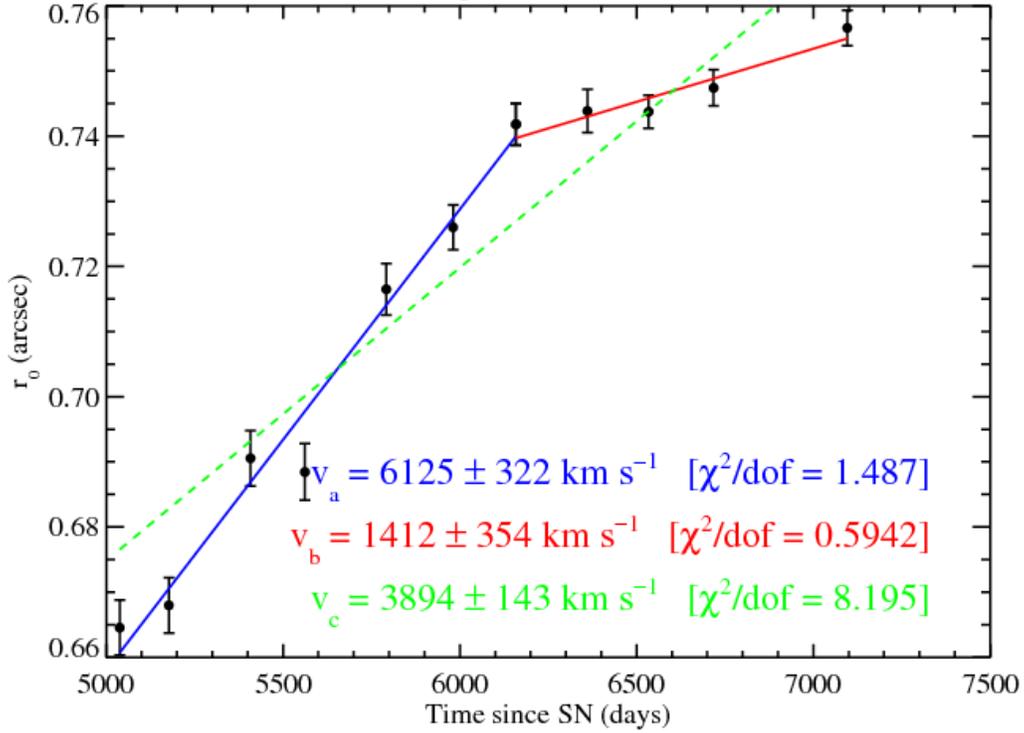}
  \caption{Radial expansion of SNR 1987A (taken from Racusin et al. in
preparation). Data taken with the gratings are excluded. Day 
4711 has also been excluded because of the low photon statistics.
}
\end{figure}

Assuming the apparent X-ray morphology of SNR 1987A (i.e., an elliptical 
torus superposed with 3$-$4 bright lobes), we model X-ray images to derive 
the best-fit radius at each epoch. The details of our image modeling 
are presented in the literature (Racusin et al. in preparation). Measured 
radii indicate that the X-ray remnant is expanding with an overall 
expansion rate of $v$ $\sim$ 3900 km s$^{-1}$ (Fig.~2), which is 
consistent with our previous estimates \cite{park04}. It is, however, 
intriguing to note that the expansion rate is significantly reduced to 
$v$ $\sim$ 1400 km s$^{-1}$ since day $\sim$ 6200 (Fig.~2). Deceleration 
of the expansion rate is in fact in good agreement with our interpretation 
of the shock reaching dense CSM all around the inner ring on days 
$\sim$6000$-$6200 \cite{park05}.

The putative neutron star has not yet become visible \cite{burrows00,
park02,park04}. If the extinction for the SNR's center were similar 
to that for the entire SNR, an upper limit of $L_{\rm X}$(2$-$10 keV) 
$\sim$ 1.5 $\times$ 10$^{34}$ ergs s$^{-1}$ has been estimated for 
an embedded point source \cite{park04}.


\section{X-Ray Spectrum}

\begin{figure}
  \includegraphics[width=\textwidth, angle=-0]{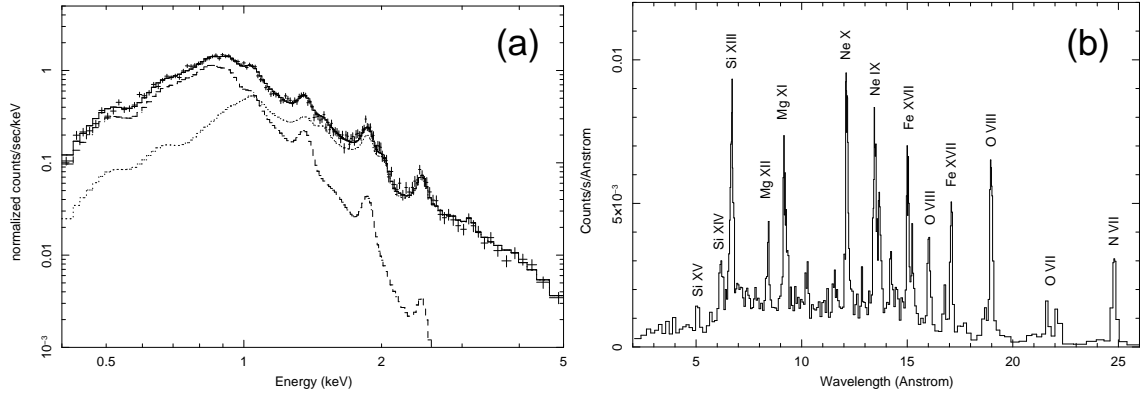}
  \caption{(a) The ACIS spectrum of SNR 1987A as of 2007-1-19. The
best-fit two-component plane shock model is overlaid. (b) The LETG
spectrum of SNR 1987A as of 2004-8 (taken from \cite{zhekov05}).
}
\end{figure}

\begin{table}
\begin{tabular}{ccccccc}
\hline
\tablehead{1}{c}{b}{Age\tablenote{Days since SN}\\(days)}
& \tablehead{1}{c}{b}{kT(soft)\\(keV)}
& \tablehead{1}{c}{b}{kT(hard)\\(keV)}
& \tablehead{1}{c}{b}{n$_{\rm e}$t(hard)\\(10$^{11}$ cm$^{-3}$ s)}  
& \tablehead{1}{c}{b}{EM(soft)\\(10$^{58}$ cm$^{-3}$)}
& \tablehead{1}{c}{b}{EM(hard)\\(10$^{58}$ cm$^{-3}$)}
& \tablehead{1}{c}{b}{$\chi$$^2$/$\nu$}\\
\hline
6914 & 0.31$^{+0.04}_{-0.02}$ & 2.21$^{+0.16}_{-0.07}$ & 
2.24$^{+0.48}_{-0.40}$ & 29.28$^{+5.86}_{-6.75}$ & 3.54$^{+0.27}_{-0.21}$ 
& 178.3/142\\
7095 & 0.29$^{+0.01}_{-0.01}$ & 2.03$^{+0.13}_{-0.12}$ & 
2.63$^{+0.62}_{-0.44}$ & 37.89$^{+1.53}_{-0.87}$ & 4.65$^{+0.30}_{-0.30}$ 
& 240.5/141 \\
7271 & 0.31$^{+0.07}_{-0.01}$ & 1.96$^{+0.09}_{-0.07}$ & 
3.63$^{+1.05}_{-0.78}$ & 40.80$^{+3.00}_{-13.80}$ & 5.61$^{+0.44}_{-0.33}$
& 183.6/142 \\
\hline
\end{tabular}
\caption{Best-Fit Parameters from the Two-Shock Model Fit of 
SNR 1987A}
\label{tab:b}
\end{table}

The X-ray spectrum of SNR 1987A is line-dominated, indicating a thermal 
origin (Fig.~3). As the shock interacts with increasing amount of 
dense CSM, multiple components of hot optically thin plasma are 
required to adequately fit the observed X-ray spectrum \cite{park04,
park06,zhekov06}. In fact, a two-temperature NEI plane shock model fits 
the observed ACIS spectrum of SNR 1987A (Fig.~3a). The soft and hard 
components characteristically represent the decelerated shock 
(by dense protrusions of the inner ring) and the fast shock propagating 
into less-dense medium, respectively. Results from two-component plane 
shock model fits of the ACIS spectrum for the latest three epochs,
which have not been published, are presented in Table~2. The foreground 
column is fixed at $N_{\rm H}$ = 2.35 $\times$ 10$^{21}$ cm$^{-2}$ 
\cite{park06}. Metal abundances are fixed at values measured by Zhekov 
et al. \cite{zhekov06}, which are generally consistent with the LMC 
abundances. Ionization timescales for the soft component ($kT$ $\sim$ 
0.3 keV) are high ($n_{\rm e}t$ $>$ 10$^{12}$ cm$^{-3}$ s), indicating 
the hot gas is in collisional ionization equilibrium due to the shock 
interaction with dense CSM. 

The high resolution dispersed spectrum obtained by the deep LETG 
observation revealed detailed X-ray emission lines from various elemental 
species (Fig.~3b, \cite{zhekov05}). The high-quality LETG spectrum 
showed that the continuous distribution of the shock temperature is
represented by two dominant components ($kT$ $\sim$ 0.5 and 2.5 keV)
\cite{zhekov06}. The LETG spectrum indicated LMC-like metal abundances 
with a moderate enhancement in N \cite{zhekov06}. X-ray line broadening 
measurements using the deep LETG observation indicated shock velocities 
of $v$ $\sim$ 300$-$1700 km s$^{-1}$ \cite{zhekov05} which are
significantly lower than that deduced from the HETG observation performed 
$\sim$5 yr earlier ($v$ $\sim$ 3400 km s$^{-1}$, \cite{michael02}). 
These results are consistent with the ACIS spectral analysis, supporting
the interpretation of the blast wave recently interacting with the entire 
inner ring.

\section{X-Ray Light curves}

\begin{table}
\begin{tabular}{cccc}
\hline
\tablehead{1}{c}{b}{Age\tablenote{Days since SN}}
& \tablehead{1}{c}{b}{f$_{\rm X}$(0.5$-$2 keV)\tablenote{Observed flux in units
of 10$^{-13}$ ergs cm$^{-2}$ s$^{-1}$}}
& \tablehead{1}{c}{b}{f$_{\rm X}$(3$-$10 keV)$^{\dagger}$}
& \tablehead{1}{c}{b}{L$_{\rm X}$(0.5$-$10 keV)\tablenote{In units of 
10$^{35}$ ergs s$^{-1}$, after corrected for $N_{\rm H}$ = 2.35 $\times$ 
10$^{21}$ cm$^{-2}$.}}\\
\hline
4711 & 1.61$\pm$0.66 & 0.84$\pm$0.57 & 1.54 \\
5038 & 2.40$\pm$0.22 & 0.92$\pm$0.21 & 2.22 \\
5176 & 2.71$\pm$0.54 & 1.22$\pm$0.41 & 2.59 \\
5407 & 3.55$\pm$0.43 & 1.20$\pm$0.44 & 3.24 \\
5561 & 4.19$\pm$0.46 & 1.49$\pm$0.64 & 3.79 \\
5791 & 5.62$\pm$0.45 & 1.82$\pm$0.46 & 5.05 \\
5980 & 6.44$\pm$0.52 & 1.95$\pm$0.62 & 5.71 \\
6157 & 7.73$\pm$0.62 & 2.38$\pm$0.57 & 6.82 \\
6359 & 11.48$\pm$0.69 & 2.40$\pm$0.60 & 9.54 \\
6533 & 16.29$\pm$0.65 & 2.80$\pm$0.73 & 13.58 \\
6716 & 19.41$\pm$0.97 & 3.26$\pm$0.68 & 16.06 \\
6914 & 21.96$\pm$1.10 & 3.45$\pm$0.69 & 17.99 \\
7095 & 25.56$\pm$1.28 & 3.84$\pm$0.77 & 20.58 \\
7271 & 29.62$\pm$1.48 & 4.41$\pm$0.88 & 23.54 \\
\hline
\end{tabular}
\caption{{\it Chandra} Flux and Luminosity of SNR 1987A}
\label{tab:c}
\end{table}
\begin{figure}
  \includegraphics[width=0.70\textwidth, angle=-0]{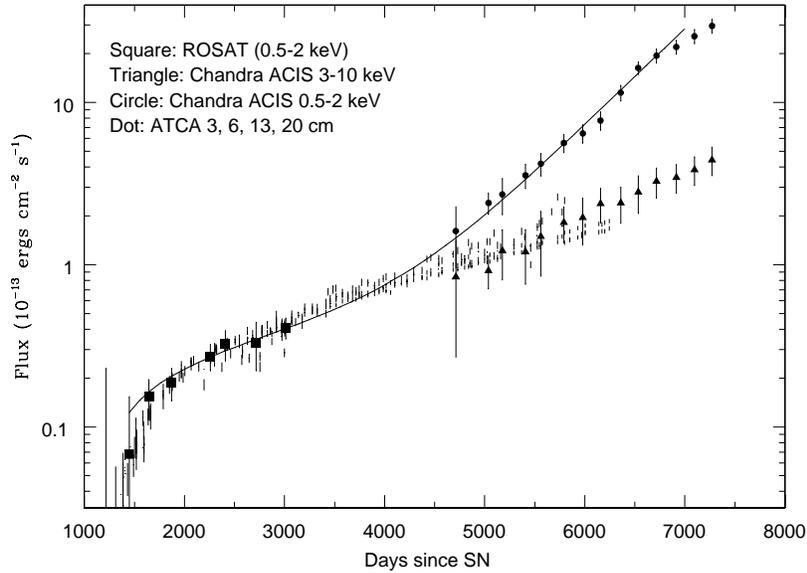}
  \caption{X-ray and radio light curves of SNR 1987A. Radio fluxes are
arbitrarily scaled. The solid line is the best-fit model by \cite{park05}}.
\end{figure}

We present the soft (0.5$-$2 keV) and hard (3$-$10 keV) band X-ray light 
curves in Table~3 and Fig.~4. We also present the {\it ROSAT} 
\cite{hasinger96} and radio\footnote{Radio data obtained with {\it 
Australian Telescope Compact Array} ({\it ATCA}) have been provided by 
L. Staveley-Smith.} light curves (Fig.~4). The soft X-ray light curve has 
been increasing nearly exponentially for the last several yr, with apparent 
``upturns'' on days $\sim$3500$-$4000 and days $\sim$6000$-$6200. These 
features were interpreted as the time when the blast wave first made  
contact with the dense protrusions, and the time when the shock
reached the main body of the inner ring \cite{park05}. The latest data 
points (days $>$ 6700) suggest that the soft X-ray flux is still rapidly 
increasing, but probably less steeply than it was for the previous 
$\sim$2 yr (Fig.~4). This latest behavior of the soft X-ray light curve 
might have implications for the details of the density structure 
of the inner ring. Periodic monitoring of the soft X-ray flux is
important to study the details of the density and abundance structures
of the inner ring. 

The hard X-ray light curve is increasing at a lower rate than the soft 
X-ray light curve (Fig.~4). This slow increase rate appears to be roughly 
consistent with the radio light curve (Fig.~4).  Hard X-rays in SNR 1987A 
might thus originate from the same synchrotron radiation as radio 
emission does. However, the morphology of hard X-ray images is no longer 
distinguishable from that of soft X-ray images \cite{park05}. The origin 
of hard X-ray emission from SNR 1987A is thus uncertain. Periodic 
monitoring of hard X-ray and radio light curves and searching for X-ray 
lines in the hard band (e.g., Fe K lines) will be important to reveal 
the origin of hard X-ray emission.

\section{The ACIS Photon Pile-Up}

Based on their {\it XMM-Newton} data analysis, Haberl et al. \cite{haberl06} 
argued that our {\it Chandra} soft X-ray light curve \cite{park05} was 
significantly contaminated by the ACIS photon pileup. They re-estimated 
the 0.5$-$2 keV band ACIS fluxes of SNR 1987A using archival {\it Chandra} 
data, and calculated pileup correction factors for the measured ACIS
fluxes. Their 0.5$-$2 keV band flux correction factors were up to
$\sim$25\%, especially for recent epochs of days $\sim$6533 and 6716. 
Thus, they claimed that the reported upturn of the soft X-ray light curve 
on days $\sim$6000$-$6200 \cite{park05} was an artifact caused by the 
photon pileup.

Haberl et al., however, misunderstood our {\it Chandra} instrument setup 
for three epochs: we used the HETG on day 4609 and a 1/8 subarray 
of the ACIS on days 6533 and 6716, while they assumed the bare ACIS on 
day 4609 and a 1/2 subarray of the ACIS on days 6533 and 6716. Their
ACIS flux corrections for these epochs were thus incorrect. 
We note that the ACIS photon pileup is not the sole contamination, and 
Haberl et al. did not consider other issues such as the charge transfer 
inefficiency and the time-dependent quantum efficiency degradation of 
the ACIS data. A moderate discrepancy ($\la$10\%) is also known between 
{\it XMM-Newton} and {\it Chandra} due to the imperfect cross-calibration 
between them. Furthermore, SNR 1987A is an extended source as observed 
with the {\it Chandra} ACIS, whereas a pointlike source was apparently 
assumed by Haberl et al. 

Considering these technical issues, we have re-analyzed the possible 
effects of ACIS photon pile-up on our {\it Chandra} observations using 
three independent methods: PIMMS/XSPEC simulations, ACIS event 
grade distribution analysis, and the ({\it modified}) standard ACIS 
photon pileup model. Our results from these three analyses agree that 
flux correction factors due to the ACIS photon pileup are roughly several 
\% or less, with the exception of day 6157 where $\sim$15\% of the soft 
X-ray flux appeared to be lost due to photon pileup (see Park et al. 
[in preparation] for the detailed results). Based on these 
results, we confirm that the scientific conclusions by Park et al. 
\cite{park05} were not affected by the ACIS photon pileup. 

\begin{theacknowledgments}
This work was supported in part by Smithsonian Astrophysical
Observatory under {\it Chandra} grant GO6-7047X.
\end{theacknowledgments}

\end{document}